\documentclass[twocolumn]{revtex4}
\usepackage{graphicx}

\begin{document}
\author{Thomas Prellberg 
\thanks{Email address: thomas.prellberg@tu-clausthal.de} 
and 
Jaros{\l}aw Krawczyk 
\thanks{Email address: krawczyk.jaroslaw@tu-clausthal.de}}
\affiliation{Institut f\"ur Theoretische Physik, 
Technische Universit\"at Clausthal,
Arnold Sommerfeld Stra\ss e 6, 
D-38678 Clausthal-Zellerfeld, Germany}
\date{\today}
\title{Flat histogram version of the pruned and enriched Rosenbluth method}
\begin{abstract}
In this letter we present a flat histogram algorithm based on 
the pruned and enriched Rosenbluth method (PERM). This algorithm 
incorporates in a straightforward manner microcanonical reweighting 
techniques, leading to ``flat histogram'' sampling in the chosen 
parameter space. As an additional benefit, our algorithm is completely 
parameter free and, hence, easy to implement.
We apply this algorithm to interacting self-avoiding walks (ISAW), the 
generic lattice model of polymer collapse.
\end{abstract}
\maketitle

Recently, there has been revived interest in flat histogram 
algorithms \cite{wang2001},
which strive to evenly sample configuration space with respect to
a chosen parameterization, e.g. microcanonical energy. 
These algorithms are particular implementations of ``umbrella sampling''
\cite{torrie1977},
in which the configuration space is sampled according to a given
probability distribution, the so-called ``umbrella''. This umbrella
distribution is generally chosen such that the whole configuration space 
of interest is accessible in one simulation. One major difficulty hereby 
is to find a suitable umbrella distribution.

There has also been an exciting development in stochastic growth algorithms,
which are based on the Rosenbluth and Rosenbluth algorithm \cite{rosenbluth1955}. If
this algorithm, which kinetically grows configurations, gets enhanced by 
cleverly chosen enrichment and pruning steps 
\cite{grassberger1997}, one obtains the pruned and enriched Rosenbluth method (PERM),
a powerful algorithm for, e.g., simulation of the polymer collapse transition.

We present in this letter a new algorithm, flatPERM, which is a combination 
of these two types of algorithms, i.e. a flat histogram version of the 
pruned and enriched Rosenbluth method.

As opposed to earlier work in this direction \cite{bachmann2003}, in which an 
iterative scheme similar to the multicanonical algorithm \cite{berg1991} was 
used, we utilize the self-tuning capabilities of PERM directly. This leads 
to a considerable simplification of the algorithm.

While flatPERM includes umbrella sampling ideas, it is stricly speaking not
a multicanonical method, as multicanonical sampling conventionally describes
a particular iterative version of adaptive umbrella sampling, in which first
an umbrella distribution is obtained iteratively and then a final simulation
is performed with a fixed umbrella distribution.

This letter is structured as follows: we first give a pedagogical
introduction to PERM, which will then allow us to introduce 
flatPERM as a seemingly trivial extension. As an application,
we present simulations of interacting self-avoiding walks (ISAW) on the
square lattice and the simple cubic lattice. We conclude with a description
of further applications.

We will consider a rather abstract setting of configurations with a certain
size $n$, which is parameterized with an additional variable $m$. In general,
one can even consider a set of variables $m_i$, but for pedagogical reasons
we shall in this paper restrict ourselves to the case of $m$ corresponding 
to an energy $E=\epsilon m$. Both $n$ and $m$ are assumed to have
non-negative integer values. Moreover, we will need the notion of 
``atmosphere'' of a configuration, which is the number $a$ of different ways 
to continue to grow this configuration, and is also a non-negative integer.

While it is useful to present the algorithm in such an abstract setting,
it may help the reader to keep the application to interacting self-avoiding 
walks (ISAW) on a regular lattice in mind. In this case, the size 
of the configuration is the number of steps of the walk, the energy is the number
of non-consecutive nearest-neighbor bonds, and the atmosphere is the number
of non-occupied sites around the endpoint of the walk. If all the sites
around the endpoint are occupied, then the atmosphere is zero and the
walk cannot be continued.

The basis of the algorithm is the Rosenbluth and Rosenbluth algorithm,
a stochastic growth algorithm in which the configurations of interest are grown 
from scratch. The growth is kinetic, which is to say that each growth
step is selected at random from all possible growth steps. Thus, if there
are $a$ possible ways of growth, one selects one of them with probability
$p=1/a$. As this number generally changes during the growth process,
different configurations are thus generated with different probabilities,
and one needs to resort to reweighting techniques.

It is advantageous to view this algorithm as approximate 
counting, in which case the weights of the configurations serve
as estimates of the number of configurations. To understand this point
of view, imagine that we were to perform a complete enumeration of the
configuration space. Doing this requires that at each growth step we
would have to choose {\em all} the possible continuations and count them
each with equal weight. If we now select {\em fewer} configurations, we 
have to change the weight of these configurations 
accordingly, in order to correct for missing out on some configurations. 
Thus, if we choose one growth step out
of $a$ possible ones, we replace $a$ configurations with equal weight by
one ``representative'' configuration with $a$-fold weight. In this way,
the weight of each grown configuration is a direct estimate of the total 
number of configurations.

Let the atmosphere $a_n=a(\varphi_n)$ be the number of distinct ways in 
which a configuration $\varphi_n$ of size $n$ can be extended. Then, the weight 
associated with a configuration of size $n$ is the product of all the 
atmospheres $a_k$ encountered while growing this 
configuration, i.e.
\begin{equation}
W=\prod_{k=0}^{n-1}a_k\;.
\end{equation}
After having started $S$ growth chains, 
an estimator $C_n^{est}$ for the total number of 
configurations $C_n$ (the 
``infinite-temperature'' partition function, in which all configurations appear
with equal weight $1$) is given by the mean over
all generated samples $\varphi_n^{(i)}$ of size $n$ with respective
weights $W_n^{(i)}$, i.e.
\begin{equation}\label{est}
C_n^{est}=\langle W\rangle_n=\frac1S\sum_iW_n^{(i)}\;.
\end{equation}
Here, the mean is taken with respect to the total number of growth 
chains $S$, and {\em not} the number of configurations which actually 
reach size $n$.  Configurations which get trapped before they 
reach size $n$ appear in this sum with weight zero.

The problem with the Rosenbluth and Rosenbluth algorithm is two-fold. 
Firstly, the weights can vary widely in magnitude, so that the mean 
can get to be dominated by very few samples with very large weight. 
Secondly, regularly occurring trapping events, i.e. generation of 
configurations with zero atmosphere, can lead to exponential 
``attrition'', i.e. exponentially strong suppression of configurations 
of large sizes.

To overcome both of these problems, enrichment and pruning steps have
been added to this algorithm, leading to the pruned and enriched
Rosenbluth method (PERM) \cite{grassberger1997}. The basic idea is that one wishes to
suppress fluctuations in the weights $W_n^{(i)}$, as these should
on average be equal to $C_n$. Therefore, if the weight is too large,
one enriches, i.e. one makes copies of the configuration and reduces
the weight accordingly. On the other hand, if the weight is too small,
one prunes probabilistically, and, in case the pruning is unsuccessful,
continues growing with appropriately increased weight. Note that $S$, and
therefore the expression (\ref{est}) for the estimate $C_n^{est}$, is
not changed by either enriching or pruning steps.

We need to specify enrichment and pruning criteria as well as the actual
enrichment and pruning processes. 
While the idea of PERM itself is straightforward, there is now a lot of 
freedom in the
precise choice of the pruning and the enrichment steps. The ``art'' of 
making PERM perform efficiently is based to a large extent on a suitable
choice of these steps. Distilling our own experience with PERM, we 
present here what we believe to be an efficient, and, most
importantly, {\em parameter free} version.

In contrast to other expositions of PERM (e.g. \cite{hsu2003}), we propose to prune or enrich
constantly, to enable larger exploration of the configuration space (the
motivation will become clear once flatPERM is introduced below). Define
$r$ as the ratio of weight and estimated number of configurations, 
$r=W_n^{(i)}/C_n^{est}$. Then we enrich if $r>1$ and prune if $r<1$.
Moreover, the actual pruning and enrichment steps are chosen such that
the weights are set as closely as possible to $C_n^{est}$ to minimize
fluctuations:
\begin{itemize}
\item  $r>1$ $\rightarrow$ enrichment step:\\ 
make $c=\min(\lfloor r\rfloor,a_n)$ distinct copies, each with weight $\frac1cW_n^{(i)}$ (as in nPERM \cite{hsu2003a}).
\item  $r<1$ $\rightarrow$ pruning step:\\
continue growing with probability $r$ and weight $C_n^{est}$ (i.e. prune with probability $1-r$).
\end{itemize}
Note that we perform pruning and enrichment {\em after} the configuration has
been included in the calculation of $C_n^{est}$ and is used for weights
during the {\em next} growth step. Initially, the estimates $C_n^{est}$
can of course be grossly wrong, as the algorithm knows nothing about the
system it is simulating. However, even if initially ``wrong'' estimates 
are used for pruning and enrichment, in all applications considered
the algorithm can be seen to converge to the true values. It is, in a sense, 
self-tuning.

At this point it is now straight-forward to change PERM to a thermal
ensemble with inverse temperature $\beta=1/k_BT$ and energy 
$E=\epsilon m$ by multiplying the weight with the appropriate 
Boltzmann factor $\exp(-\beta E)$, which leads to an estimate of 
the partition function $Z_n(\beta)$ of the form
\begin{equation}
Z_n^{est}(\beta)=\langle W\exp(-\beta E)\rangle_n\;.
\end{equation}
The pruning and enrichment procedures are changed accordingly, replacing 
$W$ by $W\exp(-\beta E)$ and $C_n^{est}$ by $Z_n^{est}(\beta)$, and using
$r=W_n^{(i)}\exp(-\beta E_m^{(i)})/Z_n^{est}(\beta)$.
(It is in this setting that PERM is usually described.)

Alternatively, however, it is also possible to consider microcanonical 
estimators for the total number $C_{n,m}$ of configurations of size $n$ 
with energy $m$ (i.e. the ``density of states'').
An appropriate estimator $C_{n,m}^{est}$ is then given by the mean over
all generated samples $\varphi_{n,m}^{(i)}$ of size $n$ and energy $m$ 
with respective weights $W_{n,m}^{(i)}$, i.e.
\begin{equation}
C_{n,m}^{est}=\langle W\rangle_{n,m}=\frac1S\sum_iW_{n,m}^{(i)}\;.
\end{equation}
Again, the mean is taken with respect to the total number of growth 
chains $S$, and {\em not} the number of configurations $S_{n,m}$ which 
actually reach a configuration of size $n$ and energy $m$. 
The pruning and enrichment procedures are also changed accordingly, 
replacing $C_n$ by $C_{n,m}$ and using
$r=W_{n,m}^{(i)}/C_{n,m}^{est}$.

At this point it is worth pointing out that the pruning and enrichment
criterion for PERM leads to a roughly constant number of samples being 
generated at each size $n$ for PERM. In fact, one can motivate the
given pruning and enrichment criteria by stipulating that one wishes to
have a roughly constant number of samples, as this leads to the algorithm
performing an unbiased random walk in the configuration size.
Correspondingly, in the flat-energy version the algorithm performs an
unbiased random walk in both size and energy of the configuration, and
we obtain a roughly constant number of samples for each size $n$ and 
energy $m$.

It is because of the fact that the number of samples is roughly constant
in each histogram entry, that this algorithm can be seen as a
``flat histogram'' algorithm, which we consequently call flatPERM. In 
hindsight in becomes clear that PERM itself can be seen as a flat histogram 
algorithm, at it creates a roughly flat histogram in size $n$. Recognizing 
this led us to the formulation of this algorithm in the first place.

At this point we return to our discussion of the pruning and enrichment
strategies. For PERM it may be more advantageous to allow for a high 
diffusivity along the size $n$. This is done by minimizing pruning
and enrichment, i.e. at the cost of allowing larger weight fluctuations.
For flatPERM on the other hand we need to achieve diffusive behavior also
with respect to the energy variable $m$. Precisely this is achieved by 
allowing for much enrichment (and thus necessarily also pruning), as each 
set of configurations enriched at size $n$ will contribute to a 
range of different energies $m$ at size $n+1$.

Even though we view flatPERM as a flat histogram algorithm, we have not
yet explicitly conditioned the pruning and enrichment with respect to
the {\em local} number of samples $S_{n,m}$ created at each size $n$ and
energy $m$. This can be done by multiplying $r$ by $S/S_{n,m}$, as this
enhances enrichment if $S_{n,m}<S$, i.e. when there are too few samples, 
and vice versa.

One general problem with PERM is that enrichment generates large
correlation between samples, so that it would be useful to replace the
number $S_{n,m}$ of samples by a number $S_{n,m}^{eff}$ of effectively
independent samples, thereby taking account of some autocorrelation time.
This can be done heuristically by considering the average number of 
independent growth steps in a configuration. If the last enrichment has
occurred at size $n_{enr}$, then a configuration of size $n$ and energy $m$
has $n_{ind}=n-n_{enr}$ independent steps. The frequency of independent
steps in a configuration is $w_{n,m}^{(i)}=n_{ind}/n$.
We thus use
\begin{equation}
r=\frac S{S_{n,m}^{eff}}\frac{W_{n,m}^{(i)}}{C_{n,m}^{est}}
\quad\mbox{with}\quad S_{n,m}^{eff}=\sum_iw_{n,m}^{(i)}
\end{equation}
for the pruning and enrichment criterion in our algorithm.

Now that one has a flat histogram method of PERM for the whole range of
energies, one can easily modify the algorithm further to sample only a 
selected range by dividing $r$ by a ``profile shape'' $f_{n,m}$, leading to
pruning whenever $f_{n,m}$ is close to zero. This is advantageous if one
only wants to explore a restricted region of parameter space.

Even though the algorithm described thus far is free of parameters, there
is a technical problem due to the fact that initial weights are grossly
wrong, which can lead to overflow problems. This can easily be overcome 
by initially restricting the maximal size of grown configurations, e.g. by
limiting the maximal size by the number of started growth chains,
\begin{equation}
n<cS\;.
\end{equation}
The relevant number of growth chains at size $n$ is thus reduced to
$S-n/c$. For ISAW we find that a choice of delay of $c\approx 10$ is sufficient.

\begin{figure}[ht]
\includegraphics[scale=0.3]{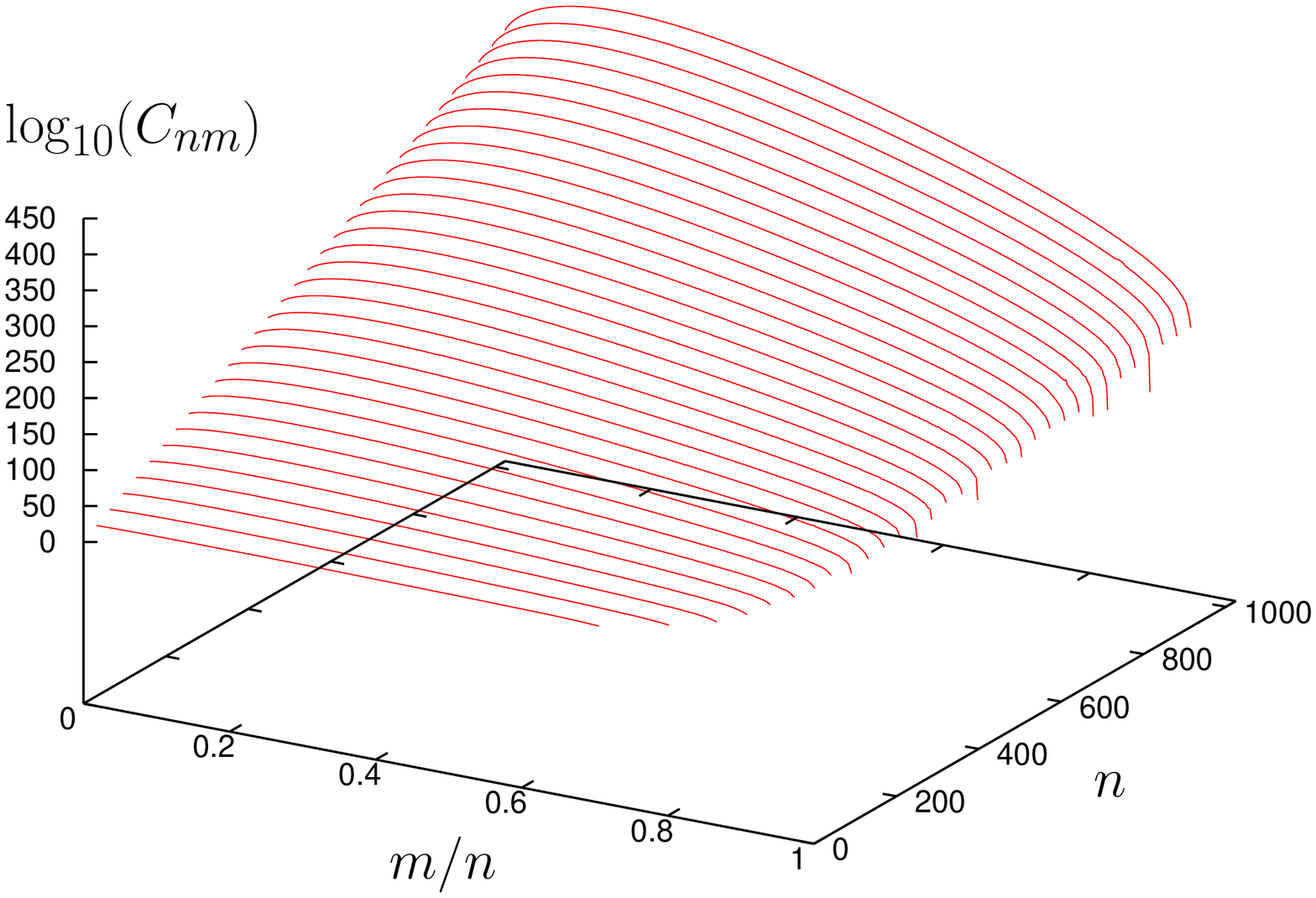}
\includegraphics[scale=0.3]{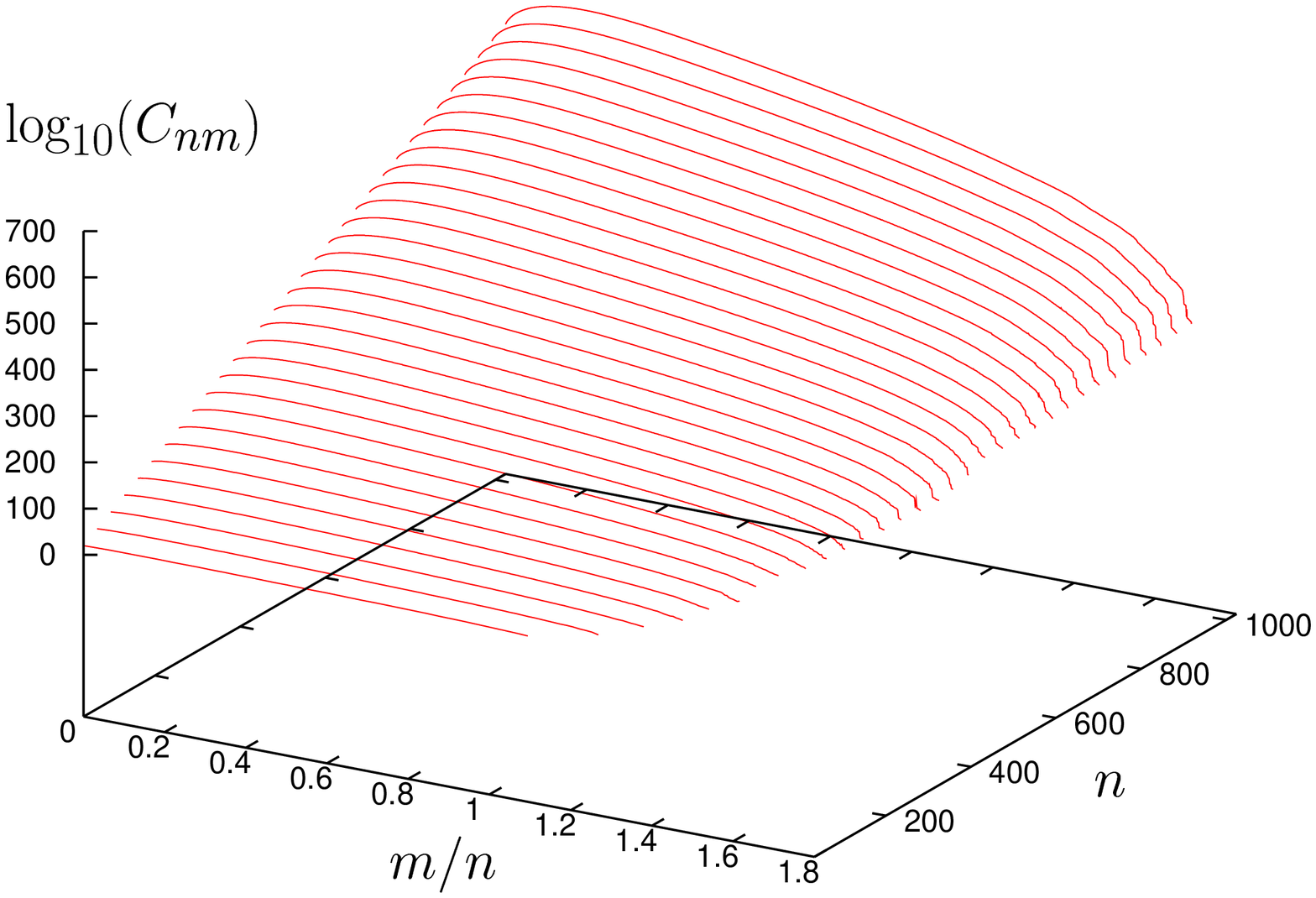}
\caption{\label{fig1} Logarithm of the number of configurations $C_{n,m}$
versus internal energy $m/n$ and length $n$ for ISAW on the square lattice
(above) and simple cubic lattice (below).}
\end{figure}

\begin{figure}[ht]
\includegraphics[scale=0.3]{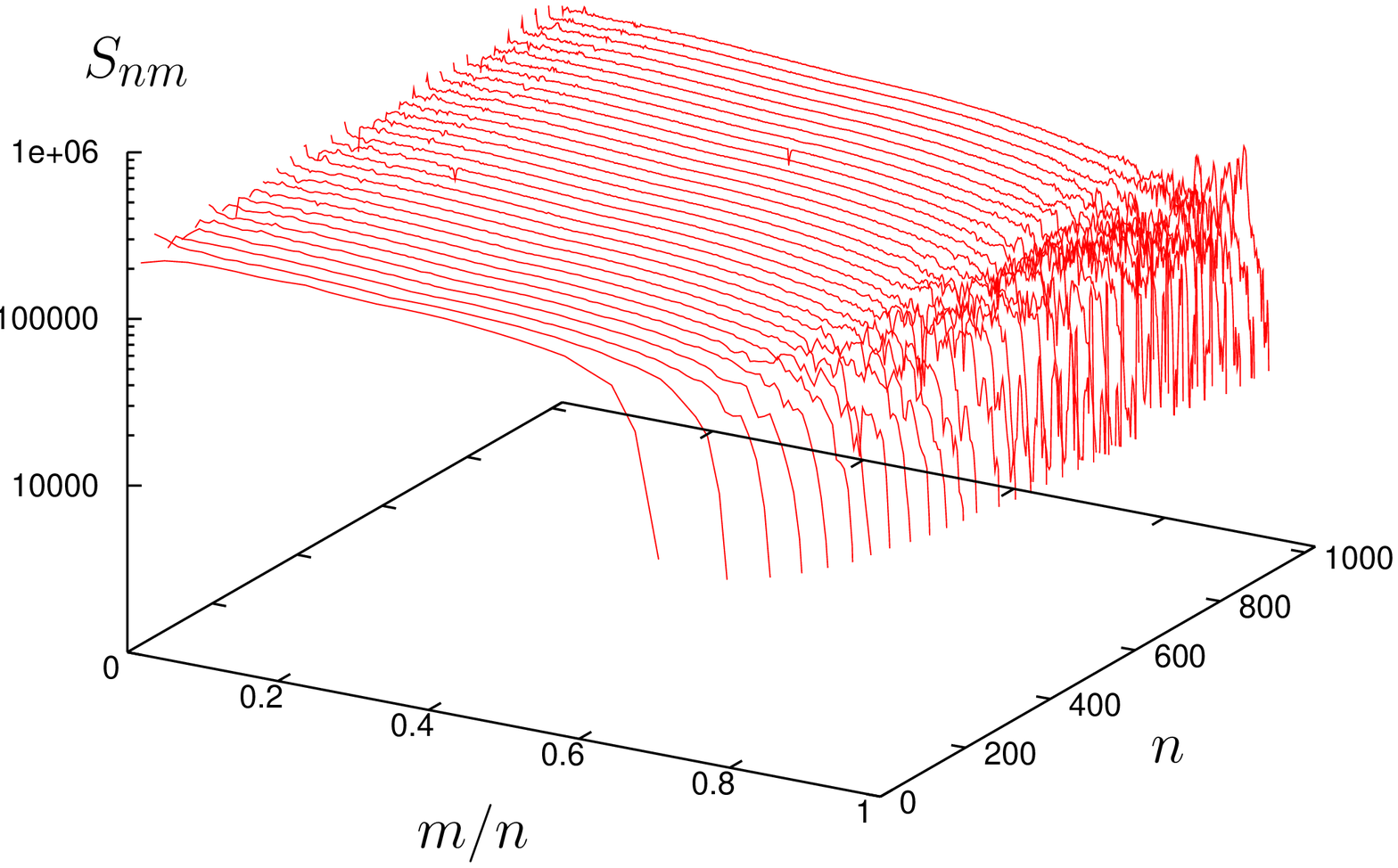}
\includegraphics[scale=0.3]{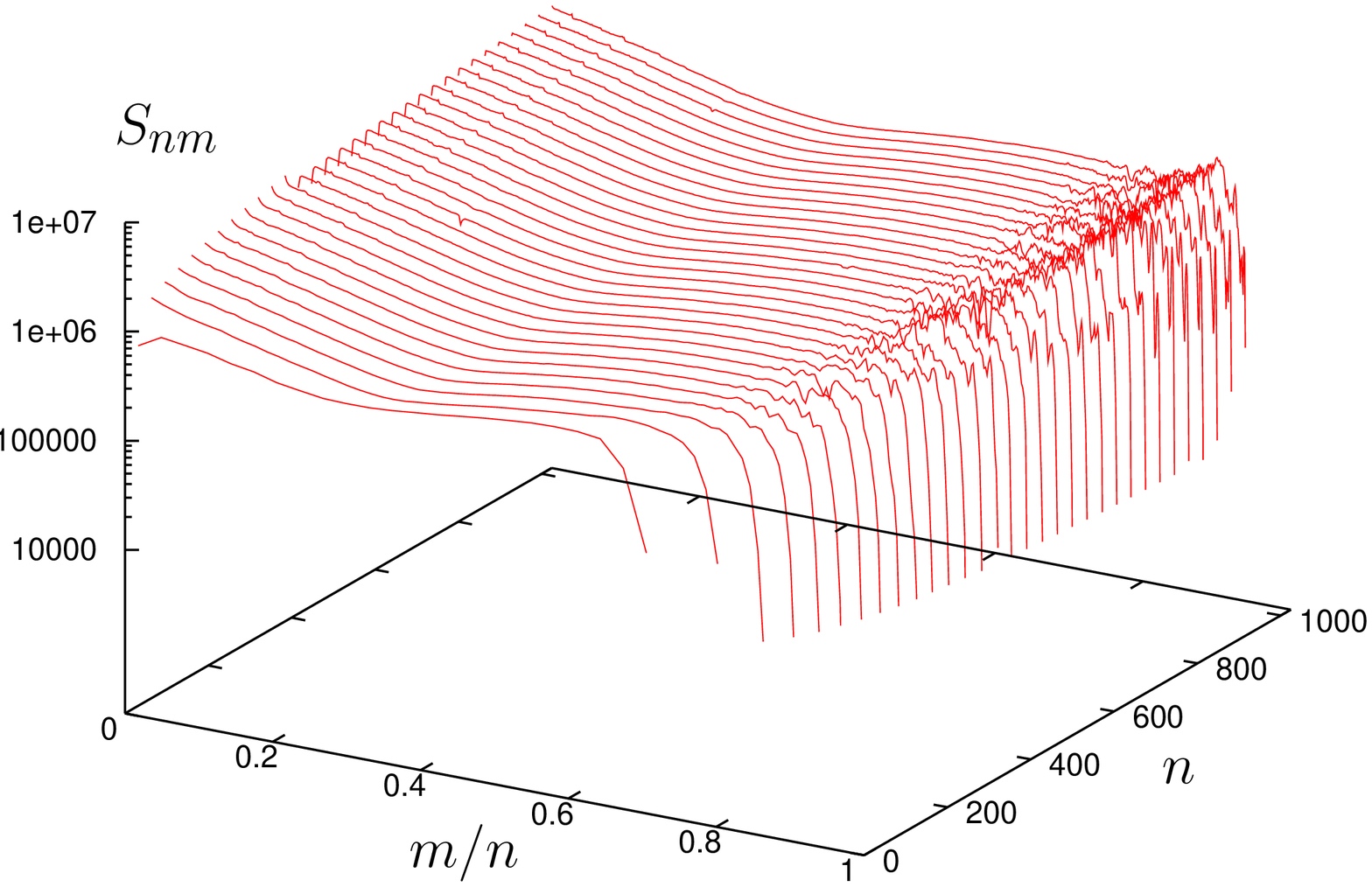}
\includegraphics[scale=0.3]{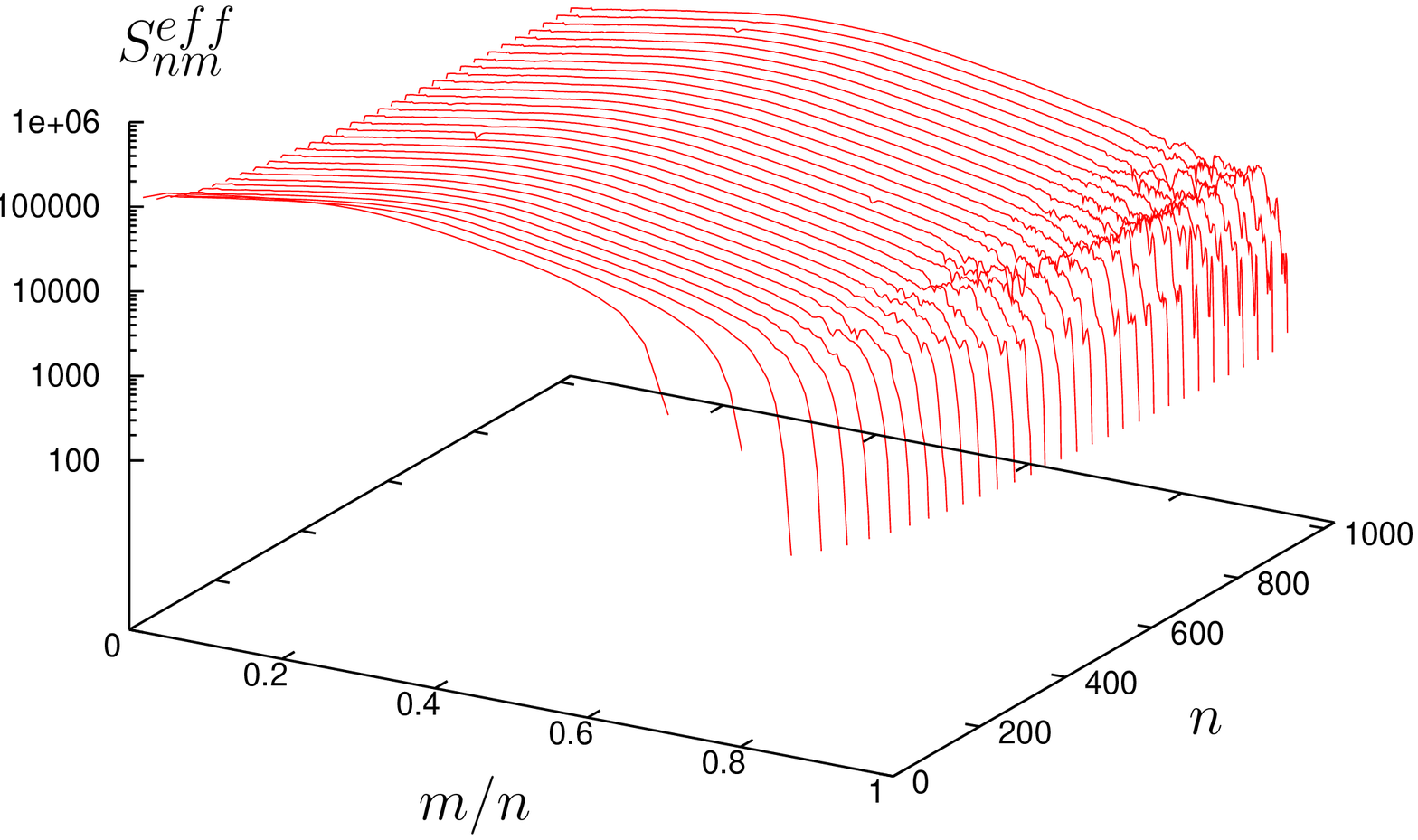}
\caption{\label{fig2} Number of generated samples versus internal 
energy $m/n$ and length $n$ for ISAW on square lattice. The top graph
shows the number of samples $S_{n,m}$ for a simulation in which
a flat histogram for $S_{n,m}$ was created, whereas the middle and bottom 
graphs show the number of samples $S_{n,m}$ and the effective number of
samples $S_{n,m}^{eff}$, respectively, for a simulation in which a 
flat histogram for $S_{n,m}^{eff}$ was created.}
\end{figure}

Averages of observables $Q$ defined on the set of configurations can now
be obtained by storing weighted sums of these observables, from which one
obtains
\begin{equation}
Q_{n,m}^{est}=\frac{\langle QW\rangle_{n,m}}{\langle W\rangle_{n,m}}=\frac{
\sum_iQ_{n,m}^{(i)}W_{n,m}^{(i)}}{\sum_iW_{n,m}^{(i)}}\;.
\end{equation}
These can then be used for subsequent evaluations. For instance, the
expectation value of $Q$ in the canonical ensemble at a given temperature
$\beta$ can now be computed via
\begin{equation}
Q_{n}^{est}(\beta)=\frac{\sum_m Q_{n,m}^{est}C_{n,m}^{est}\exp(-\beta E_m)}
{\sum_m C_{n,m}^{est}\exp(-\beta E_m)}\;.
\end{equation}

We have implemented this algorithm for  interacting self-avoiding walks on
the square and simple cubic lattice. In both two and three dimensions, we 
have simulated walks up to length $n=1024$. Figures \ref{fig1}
and \ref{fig2} clearly show the strength of the method. Figure \ref{fig1} shows
the number of configurations $C_{n,m}$, which vary over several hundred 
orders of magnitude. This range would have been inaccessible during one 
simulation with any canonical method. As an aside, we note that we had to 
rescale weights during the run to avoid overflow \cite{berg2003}. Additionally, we chose 
a delay of $c=10$ to stabilize the algorithm.

As can clearly be seen in Figure \ref{fig2}, the number 
of samples generated is roughly constant, irrespective of whether one
flattens with respect to $S_{n,m}$ (top graph) or $S_{n,m}^{eff}$ (middle and
bottom graphs). Only results for the square lattice are shown, as we find
the same behavior for the simple cubic lattice.
Using $S_{n,m}^{eff}$ for the flatness criterion leads moreover to
an increased sampling of walks with very few and very many interactions,
thus overcoming the usual difficulty of obtaining configurations with
a large number of interactions. (The largest energy state gets
repeatedly sampled in simulations in both dimensions.)

\begin{figure}[ht]
\vspace{0.3cm}
\includegraphics[scale=0.4]{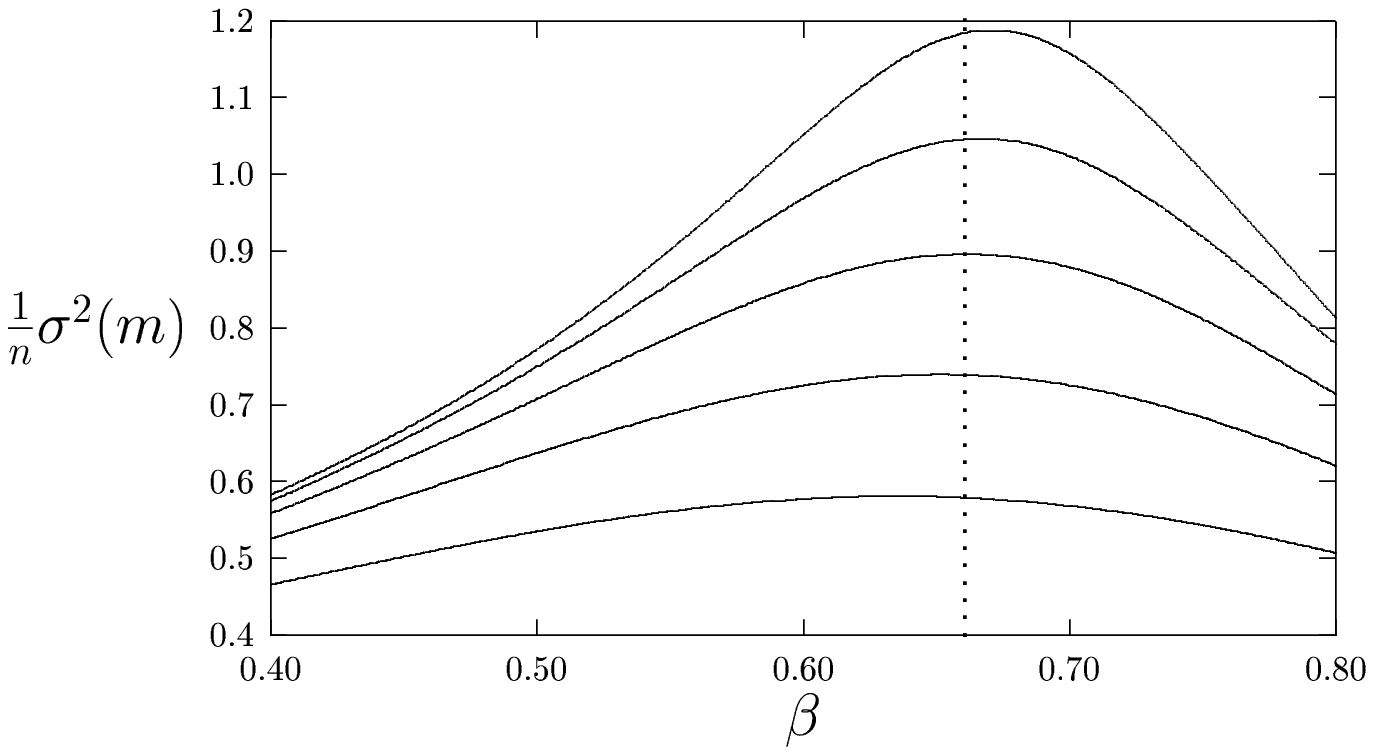}
\vspace{0.2cm}
\includegraphics[scale=0.4]{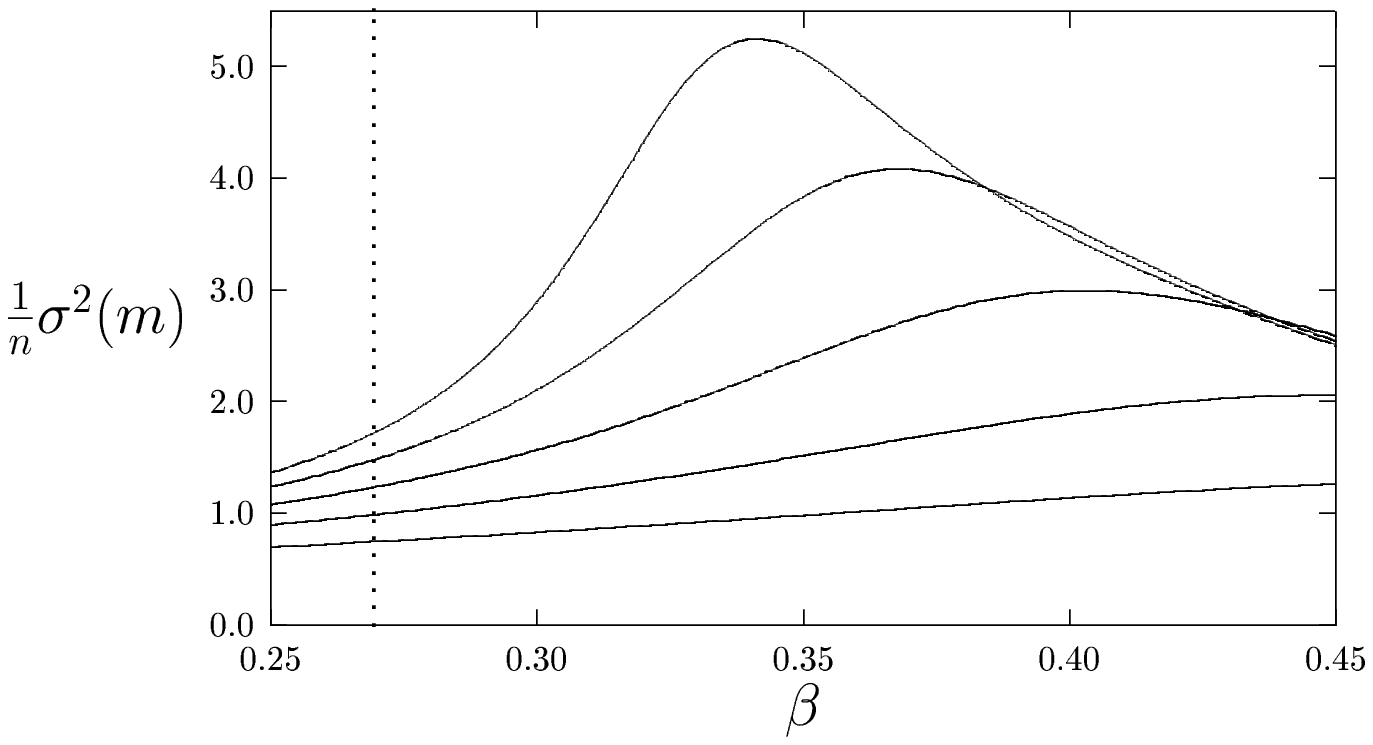}
\vspace{-0.2cm}
\caption{\label{fig3} Normalized fluctuations $\sigma^2(m)/n$ versus
inverse temperature $\beta=1/k_BT$ for ISAW on the square lattice (above) 
and the simple cubic lattice (below) at lengths 64, 128, 256, 512, and 1024.
The curves for larger lengths are more highly peaked. The vertical lines
denote the expected transition temperature at infinite length.}
\end{figure}

Once the simulations have been completed, thermodynamic quantities of
interest can easily be computed. As an example, the specific heat 
curves $C_n=k_B(\beta\epsilon)^2\sigma^2(m)/n$ near the transition 
temperature are shown for both systems in Figure \ref{fig3}.

We have also tested our algorithm on the HP model \cite{dill1985}, which is 
a toy model for proteins. It consists of a self-avoiding walk with two
types of monomers along the sites visited by the walk, which are either 
hydrophobic (type H) or polar (type P). One considers monomer-specific 
interactions, mimicking the interaction with a polar solvent such as water.
The interaction strengths are thus chosen such that HH-contacts are 
favored, e.g. by choosing $\epsilon_{HH}=-1$ and $\epsilon_{HP}=\epsilon_{PP}=0$.
The central question is to determine the density of states (and 
to find the ground state with lowest energy) for a given sequence of 
monomers. For various sequences taken from the literature we have
confirmed previous density of states calculations and obtained identical
ground state energies. The sequences we considered had $n=58$ steps
($3$ dimensions, $E_{min}=-44$) and $n=85$ steps ($2$ dimensions, $E_{min}=-53$) 
from \cite{hsu2003}, and $n=80$ steps ($3$ dimensions, $E_{min}=-98$) from \cite{frauenkron1998}.
We studied also a particularly difficult sequence with $n=88$ steps 
($3$ dimensions, $E_{min}=-73$) from \cite{beutler1996}, but the lowest energy 
we obtained was $E=-69$. However, in this case no lower energy has been 
found with any other PERM implementation either, see \cite{hsu2003}.

To summarize, we have presented a new algorithm, flatPERM, which is a 
flat histogram version of a stochastic growth algorithm. This algorithm
can in principle be applied to any statistical mechanical system for which
configurations can be grown in a well-defined manner. Next to the presented
applications of linear polymers, this algorithm can be applied to more
complicated systems, such as lattice models of branched polymers 
\cite{rechnitzer2004}. Extensions to models with two energy parameters are in 
preparation, e.g. to the problem of adsorbing interacting polymers 
\cite{owczarek2004} or an extended Domb-Joyce model of polymer collapse 
\cite{krawczyk2004}.

The authors thank Andrew Rechnitzer and Aleks Owczarek for helpful 
discussions, Peter Grassberger for pointing out an incorrect formula,
the referees for helpful comments, and the DFG for financial support.

\end{document}